\documentclass[runningheads]{llncs}
\usepackage{amsfonts,amssymb,latexsym,dsfont}
\usepackage{graphicx,epsf}
\usepackage{color}

\newcommand{\be}{\begin{equation}}
\newcommand{\ee}{\end{equation}}
\newcommand{\ba}{\begin{eqnarray}}
\newcommand{\ea}{\end{eqnarray}}
\newcommand{\ignore}[1]{}
\newcommand{\ket}[1]{\left | {#1} \right \rangle }

\newcommand{\sig}[2]{\sigma_{#1}^{#2}}
\newcommand{\la}{\lambda}
\usepackage{hyperref}
\usepackage{graphicx}

\begin{document}
\title{Towards a geometrization of quantum complexity and chaos}
\author{Davide Rattacaso\inst{1}\orcidID{0000-0001-8219-5806} \and
Patrizia Vitale\inst{1,2}\orcidID{0000-0002-5146-410X} \and
Alioscia Hamma\inst{3}\orcidID{0000-0003-0662-719X}}
\institute{ Dipartimento di Fisica Ettore Pancini, Universit\`a degli Studi di Napoli Federico II,  Via Cinthia, 80126 Fuorigrotta, Napoli NA\and
INFN-Sezione di Napoli,  Via Cinthia, 80126 Fuorigrotta, Napoli NA\and
Physics Department,  University of Massachusetts Boston,  02125, USA}

\maketitle           
\begin{abstract}
In this paper, we show how the restriction of the Quantum Geometric Tensor to manifolds of states that can be generated through local interactions provides a new tool to understand the consequences of locality in physics. After a review of a first result in this context, consisting in a geometric out-of-equilibrium extension of the quantum phase transitions, we argue the opportunity and the usefulness to exploit the Quantum Geometric Tensor to geometrize quantum chaos and complexity.
\end{abstract}

\section{Introduction}

In quantum mechanics, the state of a closed system can be represented by a point in the complex projective space $\mathds{C}P^n$, also referred to as the \emph{manifold of  pure states} $\mathcal{S}$. The tangent space of this manifold can be endowed with an Hermitian structure called \emph{Quantum Geometric Tensor} (QGT)\cite{provost}. This geometrical structure arises from the inner product on the Hilbert space and has a meaningful physical interpretation: its real part is the \emph{Fubini-Study metric} and in quantum information theory it encodes the operational distinguishability between two neighboring states of $\mathcal{S}$\cite{wooters}, while its imaginary part is called \emph{Berry's curvature} and determines the phase acquired by a state during an adiabatic cycle\cite{berry}.

The Fubini-Study metric on the manifold of pure states is homogeneous and isotropic\cite{susskindcomplexity}. This mathematical property reflects the fact that in the space $\mathcal{S}$ there are neither privileged states nor privileged evolutions. On the other side, from the physical point of view, we know that some states are privileged, in the sense that they are more easily experimentally accessible, e.g., ground states of frustration free Hamiltonians, while others can be extremely difficult to both prepare or keep from decohering in microscopic times, e.g., macroscopic superpositions (Schrodinger's cats)\cite{einselection}. The fundamental constraints on Hamiltonians and the limited resources available in terms of time, energy, and the form of interactions determine the very small region of the Hilbert space that can actually be explored by a state during its evolution\cite{poulin}. Since these constraints are at the origin of the difference between a generic quantum state and a physically \emph{accessible quantum state}, we will refer to a manifold $\mathcal{M}$ of accessible quantum states in relation to some reasonably constrained Hamiltonian, e.g., the ground states of a given family of Hamiltonians or the states that can be generated by evolution with these Hamiltonians in a finite amount of time.

Hamiltonians are constrained in the sense that not all the Hermitian operators that could in principle act on the state of a system are actually realized in nature. For example, interactions must respect some symmetry or they must observe some locality constraint. The latter request consists in the fact that the algebraic properties of accessible observables define on the Hilbert space $\mathcal{H}$ a privileged isomorphism $\mathcal{H}\cong\otimes_i^N\mathds{C}^{d_i}$\cite{TPS} such that the interactions only involve some subsets of the local Hilbert spaces $\mathds{C}^{d_i}$. These subsets can be considered as edges of a graph and therefore such a  privileged tensorization of the Hilbert space is endowed with a graph structure, and the number $N$ of these edges  represents the size of the physical system in exam. For example, the Ising Hamiltonian only involves two-spin interactions between nearest-neighbors.

One might ask whether the geometry of a manifold of quantum states can be used to investigate some physical phenomena related to the locality of the Hamiltonian exploited to access the manifold. For example, the QGT on the manifold of ground states of an Hamiltonian can reflect the capability of Hamiltonian to generate long range correlations. An insight in this direction comes from the information geometric interpretation of quantum phases by Zanardi et al.\cite{zanardiqgt,zanardiscaling,fidelitytop} where it is shown that the scaling of the first energy gap of a local Hamiltonian, and therefore its quantum phase diagram, is reflected in the scaling of the QGT on the manifold of ground states, allowing for a new information-theoretical definition of quantum phases.

In this paper, we demonstrate a formalism to describe the time evolution of quantum geometry\cite{qgtaway} defining a protocol to study the evolution of the Fubini-Study geometry on the manifold of ground states of a smooth family of local Hamiltonians, that is, the out-of-equilibrium evolution of the phase diagram. We show that quantum phases are robust after a quantum quench and the geometry of the manifold of evolving states is affected by a process of equilibration. 

Our results exemplify how the geometry of a manifold of states evolving with a local Hamiltonian can provide new answers to several problems in local dynamics. This first step serves as motivation to exploit the geometry of these manifolds as a tool to investigate many questions related to locality, like the notion of quantum complexity\cite{NielsenChuang}, the emergence of quantum phases\cite{sachdev},  equilibration processes\cite{PhysRevLett.101.190403}, operator spreading\cite{truncation}, information scrambling\cite{yoshida2,yoshida,otocANDchaos,otocANDscr} and Quantum Darwinism\cite{einselection}. Motivated by these goals, in the conclusions we outline a path towards a geometric approach to quantum complexity and chaos.

\section{Time evolved QGT and spreading of local correlations}

In this section, we demonstrate a protocol to take the manifold of ground states of an Hamiltonian out of equilibrium, obtaining time-evolving manifolds of experimentally accessible quantum states on which out-of-equilibrium quantum phases can be defined through the QGT. We show that, as it happens in the static case\cite{zanardiqgt}, the QGT encodes some properties of the system that arise from the locality of the Hamiltonian. In particular, we show that the evolution of this tensor, and therefore the geometry of the evolving manifold, is determined by the spreading of correlations between local operators. This link will be exploited in the next sections to prove some remarkable features of the quantum phases away from equilibrium.

Following the approach of the information-theoretical formulation of quantum phase transitions, we initially consider a manifold $\mathcal{M}_0$ of non-degenerate ground states of a given family of local Hamiltonians $H(\lambda)$, smooth in the control parameters $\lambda^i$. The states of this manifold can be experimentally accessed with high accuracy in a tunable synthetic quantum system at low temperature. A coordinate map on $\mathcal{M}_0$ is inherited by the control parameters $\lambda$ of the corresponding parent Hamiltonian $H(\lambda)$. To take the system out of equilibrium we introduce a family of unitary operators $U_t(\lambda):=e^{-itH^q(\lambda)}$ smooth in the control parameters $\lambda^i$. These operators describe the effect of a sudden local perturbation $H^q(\lambda)$, called quantum quench, on the corresponding state $\ket{\psi(\lambda)}$. At each time $t$ we define a manifold of out-of-equilibrium states $\mathcal{M}_t=\{|\psi_{0t}(\lambda)\rangle=U_t(\lambda)|\psi_0(\lambda)\rangle\}$, representing the evolution induced on the initial states of $\mathcal{M}_0$ by the family of quantum quenches $H^q(\lambda)$. As a consequence of the locality of $H^q$, $\mathcal{M}_t$ is a manifold of dynamically accessible quantum states. The geometrical features of this manifold can be investigated thanks to the restriction of the QGT on $\mathcal{M}_t$, that allows the definition of an out-of-equilibrium phase diagram corresponding to regions in which the QGT diverges most than linearly in the thermodynamic limit.

Given the coordinate map $x^i(\psi)$ for the manifold in exam, the QGT on this manifold can be defined via its coordinate representation
\be\label{definition}
q_{ij}(\psi):=\langle\partial_i\psi|(\mathds{1}-|\psi\rangle\langle\psi|)|\partial_j\psi\rangle.
\ee
Taking advantage of the fact that for any $t\geq0$ $\mathcal{M}_t$ is a manifold of non-degenerate ground states for the family of Hamiltonians $H(\lambda)_{t}:=U_t(\lambda)H(\lambda)U_t(\lambda)^\dag$, one can prove\cite{qgtaway} that, in a coordinate map $a^i(\psi)$ which diagonalizes the tensor, the eigenvalues $q_a$ of the QGT are bounded as follows
\begin{equation}\label{triangular}
q_0+q_1(t)-2\sqrt{q_0q_1(t)}\leq q(t)\leq q_0+q_1(t)+2\sqrt{q_0q_1(t)}
\end{equation}
where for the sake of simplicity we have eliminated the index $a$. In the last expression the term $q_0:=\sum_{n\neq0}|\langle\psi_0|\partial H|\psi_n\rangle|^2/(E_0-E_n)^2$ keeps memory of the initial geometry of the manifold, while the term $q_1(t)=\sum_{n\neq0}|\langle\psi_0|D|\psi_n\rangle|^2=\langle\psi_0|D|\psi_0\rangle^2-\langle\psi_0|D^2|\psi_0\rangle$ describes the effect of the quench on the geometry, with $D(t):=\int_0^tdt'U(t')^\dag\partial H^qU(t')$. We can deduce that the geometry of $\mathcal{M}_t$ is mainly determined by the behaviour of $q_1$. To represent $q_1$ as the sum of local correlations, we exploit the locality of the quench Hamiltonian. We can indeed represent the operator $D$ as $D=\sum_iD_i$, where $D_i:=\int_0^tdt'U(t')^\dag\partial H_i^qU(t')$\cite{qgtaway}. As a consequence $q_1=\sum_{ij}\langle D_iD_j\rangle_C$, where $\langle D_iD_j\rangle_C:=\langle \psi_0|D_iD_j|\psi_0\rangle-\langle \psi_0|D_i|\rangle\langle|D_j|\psi_0\rangle$ is the correlation function of a local operator $D_i$ which support spreads in time. Finally, making explicit the operators $D_i$, we represent the rescaled eigenvalues $q_1(\lambda,t)/N$ as
\begin{equation}\label{gcor}
q_1(\lambda,t)/N=\sum_{j}\int_0^tdt'\int_0^tdt''\langle \partial H^q_0(\lambda,t')\partial H^q_j(\lambda,t'')\rangle_C,
\end{equation}
where $N$ is the system size.
This equation made explicit the role of the spreading of local correlations in determining the geometry of  $\mathcal{M}_t$.

\section{The phase diagram away from equilibrium}

\subsection{Time evolution}
Here we show that, because of the finite velocity at which local correlations are propagated, the phase diagram, thought of as the analiticity domain of the rescaled QGT $Q/N$, is robust under a quantum quench. This means that quantum phases are stable when the system is taken out of equilibrium and that a local Hamiltonian is not capable of dynamically generating areas of fast-scaling distinguishability between states.

We suppose that the graph structure defined by the Hamiltonian on the tensorization of the Hilbert space is a lattice and we define the spacing $a$ of this lattice as the length associated with each link of the graph. In this context we say that local correlations between the operators $O_i$ and $O_j$ decay exponentially when $\langle O_iO_j\rangle_C\approx\mathcal{O}\left(e^{-\frac{a|i-j|}{\chi}}\right)$ and we call $\chi$ the correlation length of the state. Clearly the scaling in $N$ of Eq.(\ref{gcor}) directly depends on whether or not a finite correlation length in the state $\ket{\psi(\lambda,t)}$ exists.

To answer this question we exploit the Lieb-Robinson bound on the spreading of local correlations\cite{truncation}, that states that, if the evolution of the system is generated by a local Hamiltonian $H^q$ acting on a state with exponential decaying correlations, the latter  spread out at a finite velocity $v_{LR}$ called \emph{Lieb-Robinson velocity}. From this bound some important results descend, as the existence of a finite correlation length $\chi$ for non-degenerate ground states\cite{PhysRevLett.93.140402}. Generalizing these results it is possible to prove\cite{qgtaway} the following:
\begin{equation}\label{q_scaling}
|q_1(t)|/N\leq k\left[4t^2\exp{\left(\frac{2v_{LR}t}{\chi+a}   \right)}\sum_{j}\exp{\left(-\frac{d_{oj}}{\chi+a} \right)} \right]
\end{equation}
where $k$ does not depend on the time and on the size of the system. The scaling of the above equation for large $N$ depends on the behavior of the correlation length $\chi$ for the initial state $\ket{\psi_0(\lambda)}$ and, because of this bound, it is unaffected by the criticalities of the quench Hamiltonian. As a consequence the rescaled QGT, that depends on $q_1/N$ via the triangular inequality Eq.(\ref{triangular}), does not diverge in the thermodynamic limit if and only if the correlation length $\chi$ is finite. This means that the phase diagram on $\mathcal M_t$ is preserved in time and a local quench can not affect the scaling of the QGT inducing new phase transitions for some value of the the control parameters or of the time.

\subsection{Equilibration}

We are going to show that the correlation function structure of the QGT implies its equilibration in probability in the thermodynamic limit. Given a function $f(t)$ of the state of the system we say that $f(t)$ equilibrates in probability if $\lim_{N\rightarrow\infty}\sigma^2[f]\approx\mathcal{O}(e^{-N})$, where $\sigma^2[f]=\int_0^{T}(f(t)-\overline{f(t)})^2dt/T$ is the time-variance of the function over the interval of observation $T$ and $N$ is the size of the system. This behavior, that is weaker than the usual equilibration condition in which $\lim_{t\rightarrow\infty}f(t)=f_{eq}$, has been proven for the expectation value of local observables evolving with a local Hamiltonian under the non-resonance condition\cite{PhysRevLett.101.190403}. This consists  for the Hamiltonian in having a non-degenerate spectrum with also non-degenerate energy gaps. Since the degeneration is a fine-tuning condition, this request is generally satisfied. Physically this means that in a large-scale quantum system it is extremely unlikely for an experimenter to measure a local expectation value that is different from a fixed equilibration value. To extend this result to the QGT on $\mathcal{M}_t$, we show that equilibration in probability also applies to the correlation functions of local operators. In particular the following holds\cite{qgtaway}
\begin{theorem}
Consider a Hamiltonian $H^q=\sum E^q_nP^q_n$ satisfying the non-resonance condition and an observable $A(t)=e^{-itH^q}Ae^{itH^q}$ evolving under the action of $H^q$. Then the temporal variance $\sigma^2(C)$ of unequal-time correlation functions $C(t',t'') =\langle A(t')A(t'')\rangle$ is upper-bounded as $\sigma^2(C)\le \| A\|^4 Tr \overline{\rho}^2$, where $Tr(\overline{\rho}^2)$ is the purity of the completely dephased state $\overline{\rho} = \sum_n P_n^q\rho P_n^q$.
\end{theorem}

As a corollary, the same bound holds also for connected correlation functions.

This theorem is a direct extension of a previous result in literature\cite{PhysRevLett.101.190403}, and can be proven by making explicit the role of energy gaps in the time-averages involved in the definition of the time-variance. The equilibration of correlations directly influences the evolution of the QGT. Indeed from Eq.(\ref{gcor}) one can easily show that
\be
q_1(t)=\alpha t^2 + X(t)t^2
\ee
where $\alpha$ is a constant and the error $X(t)$ is a temporal variance for the involved correlation functions, that, as showed above, scales linearly in the dephased state purity. As confirmed for the Cluster-XY model in Section \ref{sec_xy}, the dephased purity $Tr(\overline{|\psi_0\rangle\langle\psi_0|}^2)$ generally decays exponentially in the system size $N$, as a consequence time fluctuations $X$ vanish very fast and $q_1$ equilibrates to $\alpha t^2$. An analogous behavior also affects the whole QGT because of the inequality Eq.(\ref{triangular}).

\subsection{Application to the Cluster-XY model phase diagram}\label{sec_xy}

Here we analyze the evolution of the QGT on the manifold of out-of-equilibrium ground states of an exactly solvable model: the Cluster-XY model\cite{hammaclusterxy}. In this way we show an example of phase-diagram evolution and how it is affected by our previous statements about scaling and equilibration in the thermodynamic limit. The Cluster-XY model consists in a spin chain evolving with the Hamiltonian
\[
 H_{CXY}(h,\lambda_x,\lambda_y)=-\sum_{i=1}^N \sig{i-1}{x}\sig{i}{z}\sig{i+1}{x}-h\sum_{i=1}^N\sig{i}{z}+\la_y\sum_{i=1}^N \sig{i}{y}\sig{i+1}{y}+\la_x\sum_{i=1}^N \sig{i}{x}\sig{i+1}{x}
\]
This model shows a very rich phase diagram on the equilibrium manifold $\mathcal{M}$.

Here we induce the evolution through a so-called \emph{orthogonal quench}: we prepare the initial manifold $\mathcal{M}_0$ as the ground states manifold of the Hamiltonians $H(\lambda)=H_{CXY}(h=0,\lambda_x,\lambda_y)$, and consider a sudden change in the control parameter of the transverse field, making the system evolve with the quench Hamiltonian $H^q(\lambda)=H_{CXY}(h=h^q,\lambda_x,\lambda_y)$. $\mathcal{M}_t$ is therefore the manifold of the states $|\Omega(\lambda,t)\rangle:=e^{-iH^q(\lambda)t}\ket{GS(H(\lambda))}$. As a consequence of the possibility to exactly diagonalize the Cluster-XY model, these states can be represented via the action of suited fermionic creation and annihilation operator on their vacuum state. This allows us to exactly calculate the overlaps between the states of the manifold and, consequently, also the metric $g$ induced from the Fubini-
Study metric of the ambient projective space. Indeed, from the squared fidelity $\mathcal F^2\equiv|\langle\Omega (\lambda',t)|\Omega (\lambda,t)\rangle\nonumber|^2$, one finds that $g_{\mu\nu}d\lambda^\mu d\lambda^\nu:=|\langle\Omega (\lambda+d\lambda,t)|\Omega (\lambda,t)\rangle\nonumber|^2$. The Berry's curvature instead can be easily shown to be zero.

The possible divergences of the metric are related to the energy gaps of the initial Hamiltonian $H(\lambda)$ and of the quench Hamiltonian $H^q(\lambda)$, anyway one can show by standard analytic techniques in \cite{qgtaway} that in the thermodynamic limit the only divergences of $g_{\mu\nu}(t)$ are the ones in $g_{\mu\nu}(0)$, determined by the disclosure of the first energy  gap of $H(\lambda)$. As a consequence the phase diagram is conserved by temporal evolution, as we have proven in the previous section. In the previous section we have also demonstrated that the equilibration properties for the QGT are determined by the purity of the dephased state $\overline{\rho}$, where $\rho=|\Omega(\lambda,t)\rangle\langle\Omega(\lambda,t)|$. Exploiting the exact diagonalization of the Cluster-XY model one can show that this purity reads $Tr(\overline{\rho}^2)= \prod_k(1-1/2\sin^2(2\chi_k))$, where $\chi_k$ is different from zero for a non-null quench. As a consequence the purity is exponentially small in $N$, determining the equilibration of the phase diagram.

\section{Conclusions and Outlook} 

The results that we have shown in this review can be considered as a first step towards the understanding of the role of the natural Hermitian structure of quantum states in determining the properties of dynamically accessible manifold of states. We extended the geometric approach to out-of-equilibrium quantum phases and, exploiting the QGT and its relation with the spreading of local correlations, we have demonstrated the robustness and equilibration of the phase diagram after a quantum quench.

Other aspects of out of equilibrium many-body physics could be investigated exploiting this geometric approach. One of the most important ones is the investigation of quantum chaotic behaviour of the Hamiltonian, which is revealed in butterfly effect and scrambling of local information\cite{yoshida2,yoshida}. A unified approach to investigate these behaviors is indeed provided by the study of out-of-time-order correlators (OTOCs) and their generalizations\cite{otocANDchaos,otocANDscr}. As a direct consequence of Eq.(\ref{gcor}), we have proven\cite{qgtaway} that the QGT is upper-bounded by a superposition of OTOCs. We can deduce that if a quench Hamiltonian generates higher distinguishability then it also generate a larger spreading of local operators. This is a first evidence of the possibility for a geometric picture of quantum chaos and scrambling. As an example, divergences in a suitable scrambling metric could allow us to identify possible quantum models of black holes thanks to the fast scrambling conjecture\cite{fastconj}.

Beyond the geometrization of chaos, the QGT could be related to quantum complexity. A metric representation of complexity has been proposed for the manifold of unitary operators\cite{Nielsen1,Nielsen2} and extended to the manifold of states\cite{susskindcomplexity}. The latter consist in modified versions of the Fubini-Study metric, in which the matrix elements associated with the infinitesimal evolutions generated only by non-local interactions are enlarged through a penalty factor. As pointed out by Nielsen\cite{Nielsen1}, such a geometric notion of complexity could play a central role in quantum computing, reducing the search for the most efficient algorithm to the search of a geodesic. Here we suggest a different approach, that does not involve an arbitrary penalty. Given a target path of states we consider the best local approximation of its generator. This time-dependent operator, that is an \emph{optimal time-dependent parent Hamiltonian}, applied to the initial point of the target path of states, generates an \emph{accessible approximation} of the latter. The functional distance between the target path and its accessible approximation measures the capability to access the target path exploiting only local interactions, so we expect that it is linked to quantum complexity. %Moreover, this distance naturally descends from the QGT, as well as the distance between two real curves only depends on the Riemannian geometry of the manifold in which the curves lie. 
Therefore, one of our next goals will be to exploit this distance to find a natural notion of complexity metric, or also \emph{accessibility metric}, on the manifold of quantum states. Our point of view on complexity geometry arises form the search for an optimal parent Hamiltonian\cite{Clark}. Since this search is directly linked to several practical goals, such as the verification of quantum devices\cite{Zoller_verification} and the quantum control of time-dependent states\cite{Gross}, the accessibility metric could provide a unified framework to understand the limits of our ability to manipulate quantum matter.


\begin{thebibliography}{8}

\bibitem{provost} Provost, J.P., Vallee, G.: Riemannian structure on manifolds of quantum states. Commun. Math. Phys. \textbf{76}, 289--301 (1980)
\bibitem{wooters} Wootters W K 1981 \emph{Phys. Rev. D} \textbf{23}, 357
\bibitem{berry} Berry, M.V.: Quantal phase factors accompanying adiabatic changes. Proc. R. Soc. Lond. A \textbf{392}(1802), 47-57 (1984)
\bibitem{susskindcomplexity} Brown, A.R., Susskind, L.: Complexity geometry of a single qubit. Phys. Rev. D \textbf{100}, 046020 (2019)
\bibitem{einselection} Zurek, W.H.: Decoherence, einselection, and the quantum origins of the classical. Rev. Mod. Phys. \textbf{75}, 715–775 (2003).
\bibitem{poulin} Poulin, D., Qarry, A., Somma, R., Verstraete, F.: Quantum Simulation of Time-Dependent Hamiltonians and the Convenient Illusion of Hilbert Space. Phys. Rev. Lett. \textbf{106}, 170501 (2011)
\bibitem{TPS} Zanardi, P., Lidar, D., Lloyd, S.: Quantum tensor product structures are observable-induced. Phys. Rev. Lett. \textbf{92}, 060402 (2004)
\bibitem{zanardiqgt} Zanardi, P., Giorda, P., Cozzini, M.: Information-Theoretic Differential Geometry of Quantum Phase Transitions. Phys. Rev. Lett. \textbf{99}, 100603 (2007)
\bibitem{zanardiscaling} Campos Venuti, L., Zanardi, P.: Quantum Critical Scaling of the Geometric Tensors. Phys. Rev. Lett. \textbf{99}, 095701 (2007)
\bibitem{fidelitytop} Abasto, D.F., Hamma, A., Zanardi,P., Fidelity analysis of topological quantum phase transitions. Phys. Rev. A \textbf{78}, 010301(R) (2008)
\bibitem{qgtaway} Rattacaso, D., Vitale, P., Hamma, A.: Quantum geometric tensor away from equilibrium. J. Phys. Comm. \textbf{4}(5), 055017 (2020)
\bibitem{NielsenChuang} Nielsen, M.A., Chuang, I.L.: Quantum Computation and Quantum Information, 10th Anniversary Edition. Cambridge University Press, New York (2010)
\bibitem{sachdev} Sachdev, S.: Quantum Phase Transitions. 2nd edn. Cambridge University Press, Cambridge (2011)
\bibitem{PhysRevLett.101.190403} Reimann, P.: Foundation of Statistical Mechanics under Experimentally Realistic Conditions. Phys. Rev. Lett. \textbf{101}, 190403 (2008)
\bibitem{truncation} Bravyi, S., Hastings, M. B., Verstraete, F.: Lieb-Robinson bounds and the generation of correlations and topological quantum order. Phys. Rev. Lett. \textbf{97}, 050401 (2006)
\bibitem{yoshida2} Hosur, P., Qi, X. L., Roberts, D. A., Yoshida, B.: Chaos in quantum channels. J. High Energy Phys. \textbf{2016}, 2 (2016)
\bibitem{yoshida} Roberts, D.A., Yoshida, B.: Chaos and complexity by design. J. High Energy Phys. \textbf{2017}, 121 (2017)
\bibitem{otocANDchaos} Hashimoto, K., Murata, K., Yoshii, R.: Out-of-time-order correlators in quantum mechanics. J. High Energy Phys. \textbf{2017}, 138 (2017)
\bibitem{otocANDscr} Gärttner, M., Hauke, P., Rey, A.M.: Relating out-of-time-order correlations to entanglement via multiple-quantum coherences. Phys. Rev. Lett. \textbf{120}, 040402 (2018)
\bibitem{PhysRevLett.93.140402} Hastings, M.B.: Locality in Quantum and Markov Dynamics on Lattices and Networks. Phys. Rev. Lett. \textbf{93}, 140402 (2004).
\bibitem{hammaclusterxy} Montes, S., Hamma, A.: Phase diagram and quench dynamics of the cluster-XY spin chain. Phys. Rev. E \textbf{86}, 021101 (2012)
\bibitem{fastconj} Lashkari, N., Stanford, D., Hastings, M., Osborne, T., Hayden, P: Towards the fast scrambling conjecture. J. High Energy Phys. \textbf{2013}, 22 (2013)
\bibitem{Nielsen1} Nielsen, M.A.: A geometric approach to quantum circuit lower bounds. arXiv:quant-ph/0502070.
\bibitem{Nielsen2}  Nielsen, M.A., Dowling, M., Gu, M., Doherty, A.C.: Quantum Computation as Geometry, Science \textbf{311}, 1133 (2006)
\bibitem{Clark} Chertkov, E., Clark, B.K.: Computational inverse method for constructing spaces of quantum models from wave functions. Phys. Rev. X \textbf{8}, 031029 (2018)
\bibitem{Zoller_verification} Carrasco, J., Elben, A., Kokail, C., Kraus, B., Zoller, P.: Theoretical and Experimental Perspectives of Quantum Verification. arXiv:2102.05927
\bibitem{Gross} Werschnik, J., Gross, E.K.U.: Quantum Optimal Control Theory. J. Phys. B: At. Mol. Opt. Phys. \textbf{40}, R175 (2007)
\end{thebibliography}
\end{document}